\documentclass[superscriptaddress,nofootinbib,notitlepage]{revtex4-1}

\usepackage{amsthm,amsfonts,amssymb,amsmath}
\usepackage{array}
\usepackage{float}
\usepackage{makecell,xcolor}
\usepackage{bm}
\usepackage{graphicx}
\usepackage{hyperref} 
\hypersetup{
    colorlinks=true,
    linkcolor=blue,
    citecolor=blue,
    filecolor=blue,
    urlcolor=blue
}

\newtheorem{theorem}{Theorem}
\newtheorem{problem}{Problem}
\newtheorem{lemma}[theorem]{Lemma}

\newcommand{\nn}{\nonumber \\}

\allowdisplaybreaks

\begin{document}
\title{Doubling the order of approximation via the randomized product formula}

\author{Chien Hung Cho}\thanks{f06222035@ntu.edu.tw}

\affiliation{Department of Physics, National Taiwan University, Taipei, Taiwan}
\affiliation{Hon Hai Quantum Computing Research Center, Taipei, Taiwan}

\author{Dominic W. Berry}\thanks{dominic.berry@mq.edu.au}
\affiliation{School of Mathematical and Physical Sciences, Macquarie University, Sydney, Australia}

\author{Min-Hsiu Hsieh}\thanks{min-hsiu.hsieh@foxconn.com}
\affiliation{Hon Hai Quantum Computing Research Center, Taipei, Taiwan}

\begin{abstract}
%Randomization has been applied to Hamiltonian simulation to improve the accuracy of product formulas. When simulation time $t$ is very large, the current randomized algorithm for the $(2k)$th-order Trotter-Suzuki formula \cite{childs2019faster} for any integer $k$ only improves the dependence on the number of summands $L$ in the Hamiltonian for the gate complexity. In this article, we achieve a better gate complexity by improving the randomized algorithm to the $(4k+1)$th-order approximation. Specifically, our $(4k+1)$th-order approximation has better gate complexity than the current product-based randomized algorithm when $t=\Omega(L)$. We provide the recipe for constructing such randomized formula and bound the error by using the mixing lemma \cite{campbell2017shorter}.
Randomization has been applied to Hamiltonian simulation in a number of ways to improve the accuracy or efficiency of product formulas.
Deterministic product formulas are often constructed in a symmetric way to provide accuracy of even order $2k$.
We show that by applying randomized corrections, it is possible to more than double the order to $4k+1$ (corresponding to a doubling of the order of the error).
In practice, applying the corrections in a quantum algorithm requires some structure to the Hamiltonian, for example the Pauli strings as are used in the simulation of quantum chemistry.
\end{abstract}
\maketitle

\section{Introduction}
Simulating quantum systems is one of the critical applications for quantum computation, which was first proposed by Richard Feynman \cite{feynman1982simulating}. When the size of the system increases, the number of parameters for describing the system grows exponentially, which leads to the difficulty of simulating such quantum systems on classical computers \cite{georgescu2014quantum}. Therefore, one could utilize the power of quantum computers to further understand the behavior of these complex quantum systems in a range of fields, for instance, quantum chemistry \cite{babbush2015chemical,poulin2014trotter,wecker2014gate,aspuru2005simulated,mcardle2020quantum}, condensed matter physics \cite{raeisi2012quantum}, and high-energy physics \cite{nachman2021quantum}.

Given the Hamiltonian $H=\sum_{j=1}^{L}H_{j}$, one of the critical tasks of quantum simulation is to construct the approximated form of the unitary, $V:=\exp(-it\sum_{j=1}^{L}H_{j})$, denoted as $U$, with elementary gates, and how to achieve this accurately and efficiently are two pivotal issues. To accurately approximate $V$, it is required that the error between $U$ and $V$ is at most $\epsilon$, while the usage of the number of qubits or elementary gates for constructing the unitary $U$ should be as small as possible. 
Often the error is described by the criterion $\Vert U-V \Vert \leq \epsilon$ in terms of the spectral norm, though for randomized formulas the diamond norm is used.

Product formulas are one of the widely-used quantum simulation methods due to their simplicity for near-term devices \cite{barends2015digital,brown2006limitations,lanyon2011universal} and have gained more attention in recent years. 
In 1996, Lloyd proposed the first-order approximation to simulate the local system with the Hamiltonian \cite{lloyd1996universal}, $H=\sum_{j=1}^{L}H_{j}$, by splitting time interval $t$ into $r$ steps, 
\begin{equation}
    \operatorname{e}^{-iHt}\approx \biggl(\operatorname{e}^{-iH_{1}t/r}\cdot \operatorname{e}^{-iH_{2}t/r}\cdots \operatorname{e}^{-iH_{L}t/r}\biggr)^{r},
\end{equation}
where it is required that $|t/r|\leq 1$. To have the higher-order approximation, Suzuki developed a method to systematically generate a product formula with $(2k)$th-order approximation \cite{suzuki1991general}. Although there are several advanced techniques having better asymptotic performance than product formulas \cite{childs2012hamiltonian,low2019hamiltonian,low2017optimal,berry2014exponential,haah2021quantum, berry2009black,berry2015simulating}, product formulas still perform well \cite{childs2018toward} when combined with information on the Hamiltonian structure \cite{childs2021theory}.

Recently, several modifications based on randomization for product formulas have been proposed. Zhang showed that product formulas with some randomized strategies are easier to implement but have the same efficiency compared to some deterministic product formulas \cite{zhang2012randomized}. Childs et al.\ proved the usefulness of randomly permuting over the summands of the Hamiltonian in the Trotter-Suzuki formula \cite{childs2019faster}, which can have better gate complexity than deterministic Trotter-Suzuki formulas. However, both the randomized Trotter from Childs et al.\ and deterministic Trotter-Suzuki formulas suffer from scaling problems when the number of summands in the Hamiltonian grows to a large number. Consequently, Campbell proposed the qDRIFT protocol and improved the dependence of the number of summands in the Hamiltonian in gate complexity \cite{campbell2019random, chen2021concentration}. Due to quadratic dependence on variable time $t$, qDRIFT performs better for a short time but gets worse after a specific critical time. Combining both advantages of qDRIFT and first-order randomized Trotter, Ouyang et al.\ proposed a randomized simulation algorithm, called SPARSTO, to simulate the Hamiltonian through sparsification \cite{ouyang2020compilation}. 

Applying randomization to further improve product formulas is therefore an important research topic for quantum simulation.
We build on existing randomized product formulas and exploit the properties of the randomized unitary channel to develop formulas with higher accuracy. In particular, we are motivated by this question:
\begin{center}
\emph{Is it possible to correct the higher-order error\\ by using randomization over the product formula ?}
\end{center}

\subsection{Overview of main results}
The answer to the above question is affirmative. In this paper, we propose a procedure to systematically construct the randomized formula with higher-order approximation, and we refer to all such formulas as \emph{modified randomized formula}. 
Our method can be applied in general, but to simulate evolution under correction terms it is most convenient if the Hamiltonian is a sum of Pauli strings.
Starting from the $(2k)$th-order approximation, our procedure allows us to increase the error order of the modified randomized formula by averaging over a correction term. This generates the modified randomized formula with $(4k+1)$th-order approximation.
As usual in this terminology, an order $2k$ approximation has error order $2k+1$, so the order of the error is being doubled from $2k+1$ to $4k+2$.

We analyze the performance of our methods using the mixing lemma \cite{campbell2017shorter}. First, we calculate the distance between each sampled unitary and the target unitary. Next, we calculate the distance between the average of the sampled unitaries and the target unitary. With the aid of the mixing lemma, when simulating the Hamiltonian $H=\sum_{j=1}^{L}H_{j}$ for time $t$, we can bound the diamond-norm distance between the randomized unitary channel proposed in this paper and the target quantum channel.
For time $t$ broken into $r$ intervals, this gives the diamond-norm distance $\mathcal{O}((tL)^{4k+2}/r^{4k+1})$ for the $(4k+1)$th-order approximation. If the diamond-norm distance may be no larger than $\epsilon$, the number of exponentials needed is $\mathcal{O}(tL^{2}(tL/\epsilon)^{\frac{1}{4k+1}})$.
The overall gate complexity of these algorithms is proportional to the number of exponentials.

In our method, the complexity of the modified randomized formula performs better than the deterministic Trotter-Suzuki formula. It also outperforms the randomized product formula proposed by Andrew Childs et al.\ when $t/\epsilon$ is large. We list the complexity of different methods in Table \ref{tab:1}. The $(4k+1)$th-order approximation provides improvements over the deterministic $(2k)$th-order Trotter-Suzuki formula with respect to all parameters of interest. As a comparison, if the term $\mathcal{O}(tL^{2}(tL/\epsilon)^{\frac{1}{4k+1}})$ dominates in the complexity of the $(2k)$th-order randomized product formula, our $(4k+1)$th-order approximation has the same complexity in this case. When $L=o((t/\epsilon)^{1+1/2k})$, the term $\mathcal{O}(tL^{2}(t/\epsilon)^{\frac{1}{2k}})$ dominates in the complexity of the $(2k)$th-order randomized product formula, and our $(4k+1)$th-order approximation is advantageous.
\begin{table}[H]
\newcommand{\tabincell}[2]{\begin{tabular}{@{}#1@{}}#2\end{tabular}} 
 \centering
\begin{center}
\begin{tabular}{ | m{19em} | m{6cm}| }
  \hline
  \thead{Method} & \thead{Number of exponentials} \\
  \hline
  \thead{$(2k)$th-order Trotter-Suzuki method} & \thead{$ \mathcal{O}(tL^{2}(tL/\epsilon)^{\frac{1}{2k}})$}\\ 
  \hline
  \thead{$(2k)$th-order randomized product formula \cite{childs2019faster}}& \thead{$\text{max}\{\mathcal{O}(tL^{2}(tL/\epsilon)^{\frac{1}{4k+1}}),\mathcal{O}(tL^{2}(t/\epsilon)^{\frac{1}{2k}}) \}$}\\
  \hline
   \thead{$(4k+1)$th-order modified randomized formula}&\thead{$\mathcal{O}(tL^{2}(tL/\epsilon)^{\frac{1}{4k+1}})$}\\
    \hline
\end{tabular}
\end{center}
\caption{\label{tab:1} The comparison of the complexity of various methods in terms of the number of exponentials.}
\end{table}

\section{Preliminaries}
In this section, we introduce some basic notations and properties of the norm, and those who are already familiar with these contents can directly skip this part.
\subsection{Norms}
Given the vector $\bm{\alpha}:=[\alpha_{1}, \alpha_{2}, \alpha_{3}, \cdots, \alpha_{L}]\in \mathbb{C}^{L}$, we define the $l_2$ norms for the vector $\bm{\alpha}$ as
\begin{equation}
\Vert \bm{\alpha}\Vert_{2}:=\sqrt{\sum_{j=1}^{L}|\alpha_j|^{2}}.
\end{equation}
If there is a matrix $A \in \mathbb{C}^{L \times L}$, the trace norm and the spectral norm are defined as
\begin{equation}
\Vert A \Vert_{1}:=\operatorname{Tr}(\sqrt{AA^{\dagger}})  \quad \quad 
\Vert A \Vert:=\max_{\bm{a}}\frac{\Vert A\bm{a}\Vert_{2}}{\Vert \bm{a}
\Vert_{2}}.
\end{equation}

Next, we define the norm for the superoperator. The diamond norm of the map $\mathcal{E}$ is defined as
\begin{equation}
\Vert \mathcal{E} \Vert_{\diamond}:=\max_{\rho: \Vert \rho\Vert_{1} \leq 1}\Vert (\mathcal{E}\otimes \mathbb{I})(\rho)\Vert_{1},
\end{equation}
where $\mathbb{I}$ acts on the same size of Hilbert space as $\mathcal{E}$.
The submultiplicativity of the diamond norm is
\begin{equation}
\Vert AB\Vert_{\diamond}\leq \Vert A\Vert_{\diamond}\Vert B\Vert_{\diamond}
\end{equation}
and this implies $\Vert A^{n}\Vert_{\diamond}\leq \Vert A\Vert_{\diamond}^{n}$.

\subsection{Trotter-Suzuki formula}
To simulate the dynamics of a system with the Hamiltonian $H=\sum_{j=1}^{L}H_{j}$, many methods have been proposed to approximate the exponentiation
 \begin{equation}\label{main_H}
     V(\lambda):= \text{exp}\biggl(\lambda \sum_{j=1}^{L}H_{j}\biggr)
 \end{equation}
 where $\lambda \in \mathbb{C}$.
 For Hamiltonian simulation $\lambda=-it$.
 The $(2k)$th-order Trotter-Suzuki formula is defined as \cite{suzuki1991general}
 \begin{equation} \label{S_{2k}}
 \begin{split}
 S_{2}(\lambda)&:=\prod_{i=1}^{L}e^{\frac{\lambda}{2} H_{i}}\prod_{i=L}^{1}e^{\frac{\lambda}{2} H_{i}}
 \\S_{2k}(\lambda)&:=S_{2k-2}(p_{k}\lambda)^{2}S_{2k-2}((1-4p_{k})\lambda)S_{2k-2}(p_{k})^{2}.
 \end{split}
 \end{equation}
 with $p_{k}:=1/(4-4^{1/(2k-1)}) $, and for each $S_{2k}$, there are $N=2 \cdot 5^{k-1}(L-1)+1$ exponential terms in it.
We could define $\mathcal{S}_{2k}$ as the quantum channel corresponding to the unitary transformation $S_{2k}$.
It is also possible to construct other symmetric product formulas \cite{yoshida1990construction}.
Our method works for these general formulas, though we will discuss the costing for the Trotter-Suzuki formula to be specific.

\subsection{Mixing lemma}
 One can approximate the target channel by using the random unitary quantum channel. The following lemma shows that the diamond-norm distance between them can be bounded by considering two factors \cite{campbell2017shorter,hastings2016turning}: one is the distance between each sampled unitary and $V$, and the other one is the distance between the average of the unitary operators and $V$.
 \begin{lemma} (Mixing lemma) \label{Mixing lemma}
    Let V be a target unitary, with an associated channel $\mathcal{V}(\rho)=V\rho V^{\dagger}$. Let ${a},{b} >0$ and $\{U_{1},U_{2},U_{3},\cdots,U_{n}\}$ be a set of unitary operators used to form a quantum channel $\mathcal{E}(\rho):=\sum_{j=1}^{n}p_{j}U_{j}\rho U_{j}^{\dagger}$ such that
    \begin{enumerate}
        \item 
        $\Vert V-U_{j} \Vert \leq a$ for all $j\in \{1,2,...,n \} $
        \item $\biggl\Vert V-\sum_{j}p_{j}U_{j} \biggr\Vert \leq b$ with some positive numbers  $p_{j}$ and $\sum_{j}p_{j}=1$.
    \end{enumerate}
    Then the error between the quantum channel $\mathcal{E}$ and $\mathcal{V}$ is bounded as
        $\Vert \mathcal{E} - \mathcal{V} \Vert_{\diamond} \leq a^{2}+2b$.
\end{lemma}

\section{Results and Construction} \label{section 2}
We first introduce the problem of interest in this paper, and give the diamond-norm distance between the modified randomized quantum channel and the target channel. Next, we provide the recipe for constructing such a modified randomized quantum channel at the end of this section.
\begin{problem}
    The Hamiltonian is in the form of $H=\sum_{j=1}^{L}H_{j}$. The problem is to present a recipe for generating the randomized product formula to the $(4k+1)$th-order approximation. In particular, we wish to construct such a modified randomized product formula with the higher-order approximation based on the order $2k$ Trotter-Suzuki formula.
\end{problem}

\begin{theorem} \label{4k+2}
Given the Hamiltonian $H=\sum_{j=1}^{L}H_{j}$, and the unitary operator $V=\exp(\lambda \sum_{j=1}^{L}H_{j})$ which corresponds to the quantum channel $\mathcal{V}:\rho \mapsto V\rho V^{\dagger}$, where $\lambda =-it/r$.
    There exists a set of unitaries $\{U_{j}\}$ and probabilities $\{p_{j} \}$ which define the random unitary quantum channel $\mathcal{E}: \rho \mapsto \sum_{j}p_{j} U_{j}\rho U_{j}^{\dagger}$ such that the error between $\mathcal{E}$ and $\mathcal{V}$ is bounded as
\begin{align}
&\Vert \mathcal{V}(\lambda)-\mathcal{E}(\lambda)\Vert_{\diamond}\leq a^2 +2b, \nn
\label{eq:thm2a}
a&= 2A, \\
\label{eq:thm2b}
        b&= 2 \frac{[(5^{k-1}+1/2)|\lambda| L\Lambda]^{4k+2}}{(4k+2)!}\exp\left((5^{k-1}+1/2)|\lambda| L\Lambda \right) +\frac{A^{2}}{2}\exp(A) +\frac {3 A^2}4 + \frac{A^3} 4 ,
\end{align}
where 
\begin{align}
    A\leq 2\frac{[(5^{k-1}+1/2)|\lambda| L\Lambda]^{2k+1}}{(2k+1)!}\exp\left((5^{k-1}+1/2)|\lambda| L\Lambda \right),
\end{align}
and $\Lambda:=\max_{j}\{\Vert H_{j}\Vert \}$.
\end{theorem}

This Theorem is obtained by using a $(2k)$th-order Trotter-Suzuki product formula, with the unitaries $U_j$ corresponding to two steps under the Trotter-Suzuki formula with a random correction in between.
The part here that is specific to the Trotter-Suzuki formula is $5^{k-1}$.
One could also use other symmetric product formulas of order $2k$ that would yield a different factor here.

Theorem \ref{4k+2} bounds the error between the quantum channel $\mathcal{E}$ and $\mathcal{V}$, which can be used to give an expression for the asymptotic error. Taking $\Lambda$ to be a constant, we set $k \in \mathbb{N}$, and $r\geq (5^{k-1}+1/2) tL\Lambda$. We then have the asymptotic error for the modified randomized formula,
\begin{equation}
\begin{split}
\biggl\Vert \mathcal{V}(-it)-\mathcal{E}^{r}(-it/r)\biggr\Vert_{\diamond}\leq \mathcal{O}\biggl(\frac{(tL)^{4k+2}}{r^{4k+1}}\biggr).
\end{split}
\end{equation}
To ensure that the simulation error is at most $\epsilon$, it suffices to use the number of segments
\begin{equation}
r_{4k+1}^{m}=\mathcal{O}\biggl(tL \biggl(\frac{tL}{\epsilon}\biggr)^{\frac{1}{4k+1}}\biggr).
\end{equation}
Multiplying by $L$ gives the order of the number of exponentials for the simulation
\begin{equation} \label{g_(4k+1)}
g_{4k+1}^{m}=\mathcal{O}\biggl(tL^{2} \biggl(\frac{tL}{\epsilon}\biggr)^{\frac{1}{4k+1}}\biggr).
\end{equation}
In contrast, for the case of the Trotter-Suzuki formula, the diamond-norm distance between $\mathcal{S}_{2k}$ and $\mathcal{V}$ is \cite{childs2019faster}
\begin{align}
    \biggl\| \mathcal{V}(-it)-\mathcal{S}_{2k}^{r}(-it/r)\biggr\|_{\diamond} \leq \mathcal{O}\biggl(\frac{(tL)^{2k+1}}{r^{2k}}\biggr).
\end{align}
To guarantee that the error is at most $\epsilon$, the number of segments $r_{2k}^{ts}$ satisfies
\begin{align}
    r_{2k}^{ts}=\mathcal{O}\biggl(tL\biggl(\frac{tL}{\epsilon}\biggr)^{\frac{1}{2k}}\biggr),
\end{align}
and this gives the order of the exponentials
\begin{align}
    g_{2k}^{ts}=\mathcal{O}\biggl(tL^{2}\biggl(\frac{tL}{\epsilon}\biggr)^{\frac{1}{2k}}\biggr).
\end{align}
When comparing to Eq.~\eqref{g_(4k+1)}, one can see that our method provides the improvement to all parameters of interest. As a comparison, for the randomized formula proposed by Andrew Childs et al.\ \cite{childs2019faster}, its number of exponentials $g_{2k}^{rand}$ is
\begin{align} \label{g_2k_rand}
    g_{2k}^{rand}=\text{max}\biggl\{\mathcal{O}\biggl(tL^{2}\biggl(\frac{tL}{\epsilon}\biggl)^{\frac{1}{4k+1}}\biggr),\mathcal{O}\biggl(tL^{2}\biggl(\frac{t}{\epsilon}\biggl)^{\frac{1}{2k}}\biggr) \biggr\}.
\end{align}
When the first term in Eq.~(\ref{g_2k_rand}) dominates, our method has the same performance as their randomized formula.
When $L=o((t/\epsilon)^{1+1/2k})$, the second term in Eq.~(\ref{g_2k_rand}) dominates, and our modified randomized formula is advantageous.

In practice, we further decompose each exponential into universal elementary gates in the quantum computer. This results in at most a constant multiplicative factor for the number of elementary gates. The exact number of gates depends on the choice of the elementary gate for the type of hardware, which is beyond the scope of our discussion. 

\subsection{Recipe for the construction of the formula} \label{recipe}
This section presents a recipe for generating the modified randomized product formula to the $(4k+1)$th-order approximation. This recipe applies to any symmetric $(2k)$th-order formula, but to be specific we restrict our discussion to Trotter-Suzuki formulas. There are two major steps for developing such formulas. First, we expand the $(2k)$th-order Trotter-Suzuki formula, $S_{2k}$, to obtain the information of the error terms. Then we employ this information to design a set of unitaries to correct the order of the distance between the target unitary and the average evolution from $(2k+1)$ to $(4k+2)$. This will yield the modified randomized product formula with $(4k+1)$th-order approximation. 

We design the average evolution, represented as $S_{4k+1}$, from the $(2k)$th-order product formula $S_{2k}$, to improve the performance of the higher-order randomized formulas. According to the mixing lemma, the accuracy of this approximation is determined by two factors: the distance between $V$ and each sampled unitary, and the distance between $V$ and the average evolution $S_{4k+1}$. The second factor is dominant in terms of $|\lambda|$ for the current higher-order randomized product formulas \cite{childs2019faster}. Therefore, we aim to construct the average evolution $S_{4k+1}(\lambda)$ such that
\begin{equation} \label{S_4k+1}
    \biggl\Vert V(\lambda)-S_{4k+1}(\lambda) \biggr\Vert =\mathcal{O}((|\lambda| \Lambda)^{4k+2}).
\end{equation} 
First, we express $S_{2k}(\lambda/2)$ as
\begin{equation} \label{S_V_D}
S_{2k}(\lambda/2)=V(\lambda/2)+D(\lambda/2),
\end{equation}
where $D(\lambda/2)$ corresponds to the difference between $S_{2k}(\lambda/2)$ and $V(\lambda/2)$. 
In particular, the following formula can approximate $V(\lambda)$ to $(4k+1)$th order by including the extra correction terms $V^{\dagger}D+DV^{\dagger}$ where both terms are for $\lambda/2$ 
\begin{equation} \label{VD}
    S_{2k}(\lambda/2)[\openone-V^{\dagger}(\lambda/2)D-DV^{\dagger}(\lambda/2)] S_{2k}(\lambda/2)=V(\lambda)+\mathcal{O}(\lambda^{4k+2}).
\end{equation}
This can be obtained from the following calculations
\begin{align}
    S_{2k}(\lambda/2) (\openone-V^{\dagger}D-DV^{\dagger}) S_{2k}(\lambda/2)
    &=S_{2k}(\lambda/2) (\openone-V^{\dagger}D)(\openone-DV^{\dagger}) S_{2k}(\lambda/2)+\mathcal{O}(\lambda^{4k+2})\nn
    &=S_{2k}(\lambda/2) (\openone+V^{\dagger}D)^{-1}(\openone+DV^{\dagger})^{-1} S_{2k}(\lambda/2)+\mathcal{O}(\lambda^{4k+2})\nn
    &=S_{2k}(\lambda/2) (V^{\dagger}S_{2k})^{-1}(S_{2k}V^{\dagger})^{-1} S_{2k}(\lambda/2)+\mathcal{O}(\lambda^{4k+2})\nn
    &=V(\lambda)+\mathcal{O}(\lambda^{4k+2}).
\end{align} 
Here all quantities with the argument omitted are for $\lambda/2$.

In fact, we can achieve the $(4k+1)$th-order approximation, when we include only the terms of $V^{\dagger}D+DV^{\dagger}$ up to $(4k+1)$th order in Eq.~\eqref{VD}.
We explicitly express the correction terms $V^{\dagger}D+DV^{\dagger}$ as
\begin{equation}
    V^{\dagger}D+DV^{\dagger}=\sum_{l\in \gamma} \frac{\lambda^{l}}{2^{l}} \mathcal{H}_{l}+ \mathcal{O}(\lambda^{4k+2}),
\end{equation}
where $\gamma$ is a set of orders used for the corrections.
The operator $\mathcal{H}_{l}$ is the linear combination of the $L_{l}$ elements in the set $\{H_{j}^{(l)}\}_{j=1}^{L_{l}}$, which is composed of the products of $l$ individual Hamiltonians from $\{H_{j}\}_{j=1}^{L}$.
Specifically, we could enumerate all the distinct terms in $\mathcal{H}_{l}$ so that
\begin{equation}
\mathcal{H}_{l}:=\sum_{j=1}^{L_{l}}\beta_{j}^{(l)}H_{j}^{(l)},
\end{equation}
where $\{\beta_{j}^{(l)}\}_{j=1}^{L_{l}}$ is the coefficient of the term $H_{j}^{(l)}$ in $\mathcal{H}_{l}$.

Due to the symmetric structure of $(V^{\dagger}{S_{2k}})^{-1}({S_{2k}}V^{\dagger})^{-1}$, the terms at orders in $\{2k+2, 2k+4,...,4k\}$ vanish simultaneously.
This can be proven from the Lemma containing Eq.~(3.5) in \cite{yoshida1990construction}.
In particular, $(V^{\dagger}{S_{2k}})^{-1}({S_{2k}}V^{\dagger})^{-1}={S_{2k}}^{\dagger}V V {S_{2k}}^{\dagger}$ satisfies time-reversibility.
This is because $S_{2k}$ is a symmetric product formula, so satisfies time-reversibility $S_{2k}(\lambda/2)S_{2k}(-\lambda/2)=\openone$.
Similarly, $V$ satisfies time-reversibility because it is the exact exponential.
Therefore, as a result of the Lemma in \cite{yoshida1990construction}, $(V^{\dagger}{S_{2k}})^{-1}({S_{2k}}V^{\dagger})^{-1}$ corresponds to an exponential containing only \emph{odd}-order terms in $\lambda$.
Moreover, it is equal to the identity up to order $2k$ (so the order $2k+1$ term is non-zero).
As a result, expanding the exponential gives the same terms up to order $4k+1$, with only odd-order terms being non-zero.
The order $4k+2$ term in the expansion of the exponential may be non-zero, because it comes from an order $2k+1$ term squared.
Note that $(V^{\dagger}S_{2k})^{-1}(S_{2k}V^{\dagger})^{-1}$ is equal to $\openone-V^{\dagger}D-DV^{\dagger}$ up to an order $4k+2$ correction, so the same result holds for orders up to $4k+1$.

Therefore, the set of orders where we need to provide corrections is $\gamma=\{2k+1,2k+3,...,4k+1\}$.
In addition, the symmetric form of $S_{2k}$ and $V^{\dagger}$ leads to the Hermitian property of the operators in $\{H_{j}^{(l)}\}_{j=1}^{L_{l}}$.
To see that, note that $(V^{\dagger}S_{2k})^{-1}(S_{2k}V^{\dagger})^{-1}$ is unitary, so is an exponential of a Hermitian operator.
Since the odd-order terms in that exponential have $\lambda$ to odd powers, $\mathcal{H}_{l}$ must be Hermitian for odd $l$.
If it happened that any terms in $\{H_{j}^{(l)}\}_{j=1}^{L_{l}}$ were not Hermitian, then we could rewrite them as ${H'_{j}}^{(l)}=(H_{j}^{(l)}+H_{j}^{(l)\dagger})/2$, and obtain Hermitian terms.
Therefore, we have the approximate form of Eq.~\eqref{VD},
\begin{equation} \label{VlD}
    S_{2k}(\lambda/2)\biggl[\openone-\sum_{l\in \gamma} \frac{\lambda^{l}}{2^{l}} \mathcal{H}_{l}\biggr] S_{2k}(\lambda/2)=S_{2k}(\lambda/2) (\openone-V^{\dagger}D-DV^{\dagger}) S_{2k}(\lambda/2)+\mathcal{O}(\lambda^{4k+2}).
\end{equation}

Although this reasoning holds for general Hamiltonians, implementing evolution under $H_{j}^{(l)}$ may be difficult.
It can be implemented efficiently in the case where the Hamiltonian is a sum of tensor products of Pauli operators, as would be suitable for quantum chemistry.
Then the correction terms are also tensor products of Pauli operators, which can be Hermitian or antiHermitian.
From the above reasoning the antiHermitian terms must cancel.

This reasoning also holds when $S_{2k}$ is replaced with an average of order $2k$ symmetric product formulae.
The reasoning to show that $S_{2k}(\lambda/2) (\openone-V^{\dagger}D-DV^{\dagger}) S_{2k}(\lambda/2)$ is equal to $V(\lambda)+\mathcal{O}(\lambda^{4k+2})$ holds unchanged.
Then the argument that $\openone-V^{\dagger}D-DV^{\dagger}$ has only odd-order Hermitian terms up to order $4k+1$ holds for any single product formula in the average.
If $D$ is computed for the average over product formulae, then the average will still be required to have only odd-order Hermitian terms.
Some later steps in our reasoning will not hold when using an average over symmetric product formulae, so we will not consider that case further.

In the following, we present a systematic procedure to construct the average evolution $S_{4k+1}$ satisfying Eq.~\eqref{S_4k+1} based on the formula in Eq.~\eqref{VlD}. 
If we could find a set of well-designed unitaries $\{U_{h}^{(l)}\}$ with some coefficients $\{\alpha_{h,l}\}$
\begin{equation}
U_{h}^{(l)}:=\exp\biggl(\alpha_{h,l}H_{h}^{(l)}\biggr),
\end{equation}
and the corresponding probabilities $\{p_{h,l}\}$ so that the average over the term $S_{2k}(\lambda/2)U_{h}^{(l)}S_{2k}(\lambda/2)$ satisfies
\begin{align} \label{SUS}
\sum_{l \in \gamma}\sum_{h=1}^{L_{l}}p_{h,l}S_{2k}(\lambda/2) U_{h}^{(l)} S_{2k}(\lambda /2)=V(\lambda)+\mathcal{O}(\lambda^{4k+2}),
\end{align}
the modified product formula in Eq.~\eqref{SUS} could yield the approximation error to order $\mathcal{O}(|\lambda|^{4k+2})$.
This expression can be satisfied provided we use the criterion for choosing $\{\alpha_{h,l}\}$ and $\{p_{h,l}\}$
\begin{equation}\label{alpha_beta_1}
p_{h,l}\alpha_{h,l}= -\biggl(\frac{\lambda}{2}\biggr)^{l}\beta_{h}^{(l)}.
\end{equation}
To show that criterion works,
\begin{align}
        \sum_{l \in \gamma}\sum_{h=1}^{L_{l}}p_{h,l}S_{2k}U_{h}^{(l)}S_{2k}
        &=  \sum_{l \in \gamma}\sum_{h=1}^{L_{l}}p_{h,l} S_{2k}\biggl(\openone + \alpha_{h,l} H_{h}^{(l)}+\mathcal{O}(\lambda^{4k+2})\biggr) S_{2k}\nn
        &=  S_{2k}\biggl(\openone-\sum_{l \in \gamma}\sum_{h=1}^{L_{l}}\biggl(\frac{\lambda}{2}\biggr)^{l}\beta_{h}^{(l)} H_{h}^{(l)} \biggr) S_{2k}+\mathcal{O}(\lambda^{4k+2}), \nn
        &=  S_{2k}\biggl(\openone-V^{\dagger}D-DV^{\dagger}\biggr) S_{2k} +\mathcal{O}(\lambda^{4k+2})\nn
        &=(V(\lambda/2))^{2}+\mathcal{O}(\lambda^{4k+2}).
\end{align}
To satisfy the criterion in Eq.~\eqref{alpha_beta_1}, we choose
\begin{align} \label{alpha}
    \alpha_{h,l} &:= -\frac{{\rm sgn}(\epsilon_{h,l})A}{\|H_{h}^{(l)}\|}\\
    p_{h,l}&:=\frac{|\epsilon_{h,l}|}{A},
\end{align}
where
\begin{align} \label{A_epsilon}
    \epsilon_{h,l} &:= \biggl(\frac{\lambda}{2}\biggr)^{l}\beta_{h}^{(l)}  \|H_{h}^{(l)}\|, \\
    A &:= \sum_{l \in \gamma}\sum_{h=1}^{L_{l}} |\epsilon_{h,l}|. \label{A}
\end{align}
Hence Eq.~\eqref{SUS} gives a recipe for constructing the formula $S_{4k+1}(\lambda)$,
and we can construct the corresponding randomized unitary quantum channel 
\begin{equation}
\mathcal{E}:\rho \to 
\sum_{l \in \gamma}\sum_{h=1}^{L_{l}}p_{h,l}~[S_{2k}(\lambda/2) U_{h}^{(l)} S_{2k}(\lambda /2)]
~\rho~
[S_{2k}(\lambda/2) U_{h}^{(l)} S_{2k}(\lambda /2)]^\dagger .
\end{equation}

\section{Proof}
We prove Theorem \ref{4k+2} in this section, and the proof consists of two major steps.  We first use the mixing lemma, Lemma~\ref{Mixing lemma}, to show that the constructed quantum channel achieves the required accuracy. Then we complete the proof by splitting it into two technical lemmas, Lemma \ref{find a_4k+2} and \ref{find b_4k+2}, whose proofs are given in Section \ref{proof lemma_a} and \ref{proof lemma_b} respectively.
\subsection{Proof of Theorem \ref{4k+2}}
When we obtain the explicit form of the formula in Section \ref{recipe},
we use Lemma \ref{Mixing lemma} to bound the accuracy of the modified randomized product formula as in Theorem \ref{4k+2}. Next, the proof is split into two parts. First, we prove the distance between each sampled unitary and the target unitary in Lemma \ref{find a_4k+2}. Then the bound of the distance between the average evolution and the target unitary is proved in Lemma \ref{find b_4k+2}.

\begin{lemma} (Find the value of $a$ for Theorem \ref{4k+2}) \label{find a_4k+2}
    For any sampled unitary in $\{U_{h}\}_{h=1}^{L_{s}}$, we have the bound
    \begin{align}
        \biggr\Vert \exp\biggl(\lambda \sum_{j=1}^{L}H_{j}\biggr)-U_{h}\biggr\Vert \leq 4 \frac{[(5^{k-1}+1/2)|\lambda| L\Lambda]^{2k+1}}{(2k+1)!}\exp\left((5^{k-1}+1/2)|\lambda| L\Lambda \right).
    \end{align}
\end{lemma}
\begin{lemma} (Find the value of $b$ for Theorem \ref{4k+2}) \label{find b_4k+2}
    The distance between $S_{4k+1}$ and $V$ is bounded as 
\begin{align}
    \biggl \|S_{4k+1}(\lambda) -V(\lambda)\biggr\| 
    &\leq 2 \frac{[(5^{k-1}+1/2)|\lambda| L\Lambda]^{4k+2}}{(4k+2)!}\exp\left((5^{k-1}+1/2)|\lambda| L\Lambda \right)\nn
     &\quad +\frac{A^{2}}{2}\exp(A) +3\left\|D \right\|^2+ 2\left\|D \right\|^3,
\end{align}
where 
\begin{align}
      \|D\| \leq \frac{(5^{k-1}|\lambda| L\Lambda)^{2k+1}}{(2k+1)!}\exp\left(5^{k-1}|\lambda| L\Lambda \right)
    +\frac{(|\lambda| L\Lambda/2)^{2k+1}}{(2k+1)!}\exp\left(|\lambda| L\Lambda/2 \right)
\end{align}
and 
\begin{align}
    A \le 2\frac{[(5^{k-1}+1/2)|\lambda| L\Lambda]^{2k+1}}{(2k+1)!}\exp\left((5^{k-1}+1/2)|\lambda| L\Lambda \right).
\end{align}
\end{lemma}

Given these Lemmas, we can prove Theorem \ref{4k+2} as follows.
\begin{proof}
Using the result in Lemma \ref{find a_4k+2}, and the upper bound on $A$, the equation $\Vert V-U_{j} \Vert \leq a$ in Lemma \ref{Mixing lemma} can be satisfied with $a$ as in \eqref{eq:thm2a}.
Then, using Lemma \ref{find b_4k+2}, the condition $\biggl\Vert V-\sum_{j}p_{j}U_{j} \biggr\Vert \leq b$ in Lemma \ref{Mixing lemma} can be satisfied with $b$ as in \eqref{eq:thm2b}.
There we have replaced $\|D\|$ with $A/2$ for simplicity, because $\|D\|\le A/2$.
Therefore, we can use Lemma \ref{Mixing lemma} to provide the bound $\Vert \mathcal{V}(\lambda)-\mathcal{E}(\lambda)\Vert_{\diamond}\leq a^2 +2b$ required for Theorem \ref{4k+2}.
\end{proof}

\subsection{Proof of Lemma \ref{find a_4k+2} } \label{proof lemma_a}
There are two major steps in this proof. We first explicitly express $V(\lambda)-U_{h}(\lambda)$ as the summation of three parts, where $U_{h}$ is the sampled unitary in $\{U_{h}\}_{h=1}^{L_{s}}$. Then we individually calculate the norm of these three parts with the aid of Lemma \ref{bound_A_D}. When we have their norms, we complete the proof by using triangle inequality.
\begin{proof}
For the sampled unitary in $\{U_{h}\}_{h=1}^{L_{s}}$, one of the sampled unitaries is written as
\begin{equation}
    S_{2k}(\lambda/2)U_{h}^{(l)}S_{2k}(\lambda/2).
\end{equation}
Next, we consider the distance between the target unitary $V$ and $S_{2k}U_{h}^{(l)}S_{2k}$
\begin{align} \label{norm_V_sampledUnitary}
    \biggl\Vert \exp\biggl(\lambda \sum_{j=1}^L H_j \biggr)-  S_{2k} U_{h}^{(l)} S_{2k} \biggr\Vert 
    &=\biggl\Vert \exp\biggl(\frac{\lambda}2 \sum_{j=1}^L H_j \biggr)\exp\biggl(\frac{\lambda}2 \sum_{j=1}^L H_j \biggr) -  S_{2k} U_{h}^{(l)} S_{2k} \biggr\Vert \nn
    &\le \biggl\Vert \biggl[\exp\biggl(\frac{\lambda}2 \sum_{j=1}^L H_j \biggr)-  S_{2k}\biggr]\exp\biggl(\frac{\lambda}2 \sum_{j=1}^L H_j \biggr)  \biggr\Vert
    + \biggl\Vert S_{2k}\left[\openone - U_{h}^{(l)} \right]\exp\biggl(\frac{\lambda}2 \sum_{j=1}^L H_j \biggr)  \biggr\Vert\nn
    & \quad + \biggl\Vert S_{2k} U_{h}^{(l)}\biggl[\exp\biggl(\frac{\lambda}2 \sum_{j=1}^L H_j \biggr)-  S_{2k}\biggr]\biggr\Vert\nn
    & \le 2 \biggl\Vert \exp\biggl(\frac{\lambda}2 \sum_{j=1}^L H_j \biggr)-  S_{2k}  \biggr\Vert + \biggl\Vert \openone - U_{h}^{(l)} \biggr\Vert .
\end{align}
Using the expression for $U_{h}^{(l)}$ above, we have
\begin{align}
    \biggl\Vert \openone - U_{h}^{(l)} \biggr\Vert &= \biggl\Vert \openone - \exp\left( \alpha_{h,l} H_h^{(l)}\right) \biggr\Vert \nn
    &\le \biggl\Vert  \alpha_{h,l} H_h^{(l)} \biggr\Vert \nn
    &= A,
\end{align}
using the expressions for $\alpha_{h,l}$ and $A$ in \eqref{alpha} and \eqref{A}.
We are considering the case where $\lambda$ is imaginary so $\alpha_{h,l}$ is as well, which gives the second line above.
Thus this error is equal to double the error of $S_{2k}$ on $\lambda/2$ plus $A$,
\begin{align}
    \biggl\Vert \exp\biggl(\lambda \sum_{j=1}^L H_j \biggr)-  S_{2k} U_{h}^{(l)} S_{2k} \biggr\Vert \leq 2\|D\| +A.
\end{align}

From the results in Lemma \ref{bound_A_D} in the Appendix, we can bound the quantities $\| D \|$ and $A$ as
\begin{align}\label{eq:s2kerror}
    \| D \| &= \biggl\Vert \exp\biggl(\frac{\lambda}2 \sum_{j=1}^L H_j \biggr)-  S_{2k}  \biggr\Vert
    \nn
    & \le \frac{(5^{k-1}|\lambda| L\Lambda)^{2k+1}}{(2k+1)!}\exp\left(5^{k-1}|\lambda| L\Lambda \right)
    +\frac{(|\lambda| L\Lambda/2)^{2k+1}}{(2k+1)!}\exp\left(|\lambda| L\Lambda/2 \right),
\end{align}
and
\begin{align}
    A \le 2 \frac{[(5^{k-1}+1/2)|\lambda| L\Lambda]^{2k+1}}{(2k+1)!}\exp\left((5^{k-1}+1/2)|\lambda| L\Lambda \right).
\end{align}
This gives the bound
\begin{align} \label{bound_a}
    \biggl\Vert \exp\biggl(\lambda \sum_{i=1}^L H_i \biggr)-  S_{2k} U_{h}^{(l)} S_{2k} \biggr\Vert
    &\leq 2\frac{(5^{k-1}|\lambda| L\Lambda)^{2k+1}}{(2k+1)!}\exp\left(5^{k-1}|\lambda| L\Lambda \right)
    +2\frac{(|\lambda| L\Lambda/2)^{2k+1}}{(2k+1)!}\exp\left(|\lambda| L\Lambda/2 \right)\nn
    &\quad \quad+ 2 \frac{[(5^{k-1}+1/2)|\lambda| L\Lambda]^{2k+1}}{(2k+1)!}\exp\left((5^{k-1}+1/2)|\lambda| L\Lambda \right) \nn
    &\leq  4 \frac{[(5^{k-1}+1/2)|\lambda| L\Lambda]^{2k+1}}{(2k+1)!}\exp\left((5^{k-1}+1/2)|\lambda| L\Lambda \right).
\end{align}
\end{proof}

\subsection{Proof of Lemma \ref{find b_4k+2} } \label{proof lemma_b}
There are two steps in the proof of Lemma \ref{find b_4k+2}. First of all, we expand $S_{4k+1}(\lambda)$, and this gives us the difference between $S_{4k+1}(\lambda)$ and $V(\lambda)$. Next, we can bound the distance between $S_{4k+1}(\lambda)$ and $V(\lambda)$ by using the triangle inequality. After we have the bound of these individual terms, we prove Lemma \ref{find b_4k+2}.
\begin{proof}
We explicitly expand $S_{4k+1}$ as
\begin{align}
        \sum_{l \in \gamma}\sum_{h=1}^{L_{l}}p_{h,l}S_{2k}U_{h}^{(l)}S_{2k}
        &=  \sum_{l \in \gamma}\sum_{h=1}^{L_{l}}\frac {|\epsilon_{h,l}|}{A}S_{2k} \biggl(\openone+ \alpha_{h,l} H_{h}^{(l)}+\sum_{j=2}^{\infty}\frac{1}{j!}(\alpha_{h,l} H_{h}^{(l)})^{j}\biggr)S_{2k} \nn
        &= S_{2k}\biggl(\openone-\sum_{l \in \gamma}\sum_{h=1}^{L_{l}}\biggl(\frac{\lambda}{2}\biggr)^{l}\beta_{h}^{(l)} H_{h}^{(l)} \biggr)S_{2k} + S_{2k}\biggl(\sum_{l\in \gamma}\sum_{h=1}^{L_{l}} \frac {|\epsilon_{h,l}|}{A}\sum_{j=2}^{\infty}\frac{1}{j!}(\alpha_{h,l} H_{h}^{(l)})^{j}\biggr)S_{2k}\nn
        &=S_{2k}(\openone-V^{\dagger}D-DV^{\dagger})S_{2k} + S_{2k}\biggl(\sum_{l\in \gamma}\sum_{h=1}^{L_{l}} \frac {|\epsilon_{h,l}|}{A}\sum_{j=2}^{\infty}\frac{1}{j!}(\alpha_{h,l} H_{h}^{(l)})^{j}\biggr)S_{2k}\nn
        & \quad \quad + S_{2k}[R_{4k+1}(V^{\dagger}D+DV^{\dagger})]S_{2k} .
\end{align}
Next, note that the error in $S_{2k}(I-V^{\dagger}D-DV^{\dagger})S_{2k}$ can be bounded as
\begin{align}
   &\left\| (\openone-V^{\dagger}D-DV^{\dagger}) - (\openone+V^{\dagger}D)^{-1}(\openone+DV^{\dagger})^{-1} \right\| \nn
   &=\left\| (\openone+V^{\dagger}D)(\openone-V^{\dagger}D-DV^{\dagger})(\openone+DV^{\dagger}) - \openone \right\| \nn
   &=\left\| V^{\dagger}DV^{\dagger}D+V^{\dagger}DDV^{\dagger} +DV^{\dagger}DV^{\dagger} +V^{\dagger}DV^{\dagger}DDV^{\dagger}-V^{\dagger}DDV^{\dagger}DV^{\dagger} \right\| \nn
   &\le 3\left\|D \right\|^2 + 2\left\|D \right\|^3 .
\end{align}
Now we use the triangle bound to have
\begin{align}
     \biggl \|\sum_{l \in \gamma}\sum_{h=1}^{L_{l}}p_{h,l}S_{2k}U_{h}^{(l)}S_{2k} -V^{2}(\lambda/2)\biggr\| 
     &\leq    \biggl \|S_{2k}\biggl(\sum_{l\in \gamma}\sum_{h=1}^{L_{l}} \frac {|\epsilon_{h,l}|}{A}\sum_{j=2}^{\infty}\frac{1}{j!}(\alpha_{h,l} H_{h}^{(l)})^{j}\biggr)S_{2k}\biggr \|  \label{VS_1}\\
        & \quad \quad + \biggl\|S_{2k}[R_{4k+1}(V^{\dagger}D+DV^{\dagger})]S_{2k} \biggr\| +3\left\|D \right\|^2+ 2\left\|D \right\|^3 \label{VS_2} . %\biggl\| S_{2k}V^{\dagger}D DV^{\dagger}S_{2k} \biggr\| 
\end{align}

Next, we bound each norm individually.
Among Eq.~\eqref{VS_1}, we use some standard properties of norms, the definitions in Eq.~\eqref{alpha}, Eq.~\eqref{A_epsilon}, and Eq.~\eqref{A} to have
\begin{align}
    \biggl\Vert S_{2k}\biggl(\sum_{l\in \gamma}\sum_{h=1}^{L_{l}} p_{h,l}\sum_{j=2}^{\infty}\frac{1}{j!}(\alpha_{h,l} H_{h}^{(l)})^{j}\biggr)S_{2k}\biggl\Vert 
    &\leq \Vert S_{2k} \Vert\cdot \biggl\Vert \sum_{l\in \gamma}\sum_{h=1}^{L_{l}}p_{h,l}\sum_{j=2}^{\infty}\frac{1}{j!}(\alpha_{h,l} H_{h}^{(l)})^{j} \biggl\Vert \cdot \Vert S_{2k} \Vert \nn
    &\leq \sum_{l\in \gamma}\sum_{h=1}^{L_{l}}p_{h,l} \sum_{j=2}^{\infty} \biggl\Vert \frac{1}{j!}(\alpha_{h,l} H_{h}^{(l)})^{j} \biggl\Vert \nn
    &= \sum_{j=2}^{\infty}\frac{A^{j}}{j!} \nn
    &\leq \frac{A^{2}}{2}\exp(A).
\end{align}
In the second-last line we have used the fact that the sum over probabilities is equal to 1.

For the norm in Eq.~\eqref{VS_2}, using some basic properties of norms we have
\begin{align}
     \biggl\|S_{2k}[R_{4k+1}(V^{\dagger}D+DV^{\dagger})]S_{2k} \biggr\| 
     &\leq \|S_{2k}\| \cdot \biggl\|R_{4k+1}(V^{\dagger}D+DV^{\dagger}) \biggl\| \cdot \|S_{2k} \| 
     \leq \biggl\|R_{4k+1}(V^{\dagger}D+DV^{\dagger}) \biggl\|.
\end{align}
This can be bounded using Eq.~\eqref{eq:orderserrorVD}, and summing from $s=4k+2$ to infinity with $\lambda$ replaced with $\lambda/2$ for the half-interval to give
\begin{align}
   2\sum_{s=4k+2}^\infty \frac{[(5^{k-1}+1/2)|\lambda| L\Lambda]^s}{s!} &\le 
    2 \frac{[(5^{k-1}+1/2)|\lambda| L\Lambda]^{4k+2}}{(4k+2)!}\exp\left((5^{k-1}+1/2)|\lambda| L\Lambda \right).
\end{align}

As a result, we have
\begin{align}
    &\biggl \|\sum_{l \in \gamma}\sum_{h=1}^{L_{l}}p_{h,l}S_{2k}U_{h}^{(l)}S_{2k} -V^{2}(\lambda/2)\biggr\| \nn
    &\leq 2 \frac{[(5^{k-1}+1/2)|\lambda| L\Lambda]^{4k+2}}{(4k+2)!}\exp\left((5^{k-1}+1/2)|\lambda| L\Lambda \right)
      +\frac{A^{2}}{2}\exp(A) +3\left\|D \right\|^2+ 2\left\|D \right\|^3, 
\end{align}
where we bound $\| D\|$ and $A$ in Lemma \ref{bound_A_D}.
\end{proof}

\section*{Acknowledgment}
DWB worked on this project under a sponsored research agreement with Google Quantum AI.
DWB is also supported by Australian Research Council Discovery Project DP210101367.
CHC thanks Ching Ray Chang for the kind support and comments.
\bibliographystyle{apsrev4-1}
\bibliography{bibliography_0}

\appendix
\section{Proofs of operator bounds}
\begin{lemma} \label{bound_A_D}
    Defining $A$ and $D$ as in Eq.~\eqref{A} and Eq.~\eqref{S_V_D}, we have the upper bounds
    \begin{align}
        \|D\| \leq \frac{(5^{k-1}|\lambda| L\Lambda)^{2k+1}}{(2k+1)!}\exp\left(5^{k-1}|\lambda| L\Lambda \right)
    +\frac{(|\lambda| L\Lambda/2)^{2k+1}}{(2k+1)!}\exp\left(|\lambda| L\Lambda/2 \right),
    \end{align}
    and
    \begin{align}
         A \le 2\frac{[(5^{k-1}+1/2)|\lambda| L\Lambda]^{2k+1}}{(2k+1)!}\exp\left((5^{k-1}+1/2)|\lambda| L\Lambda \right).
    \end{align}
\end{lemma}
\begin{proof}
Using the approach in \cite{berry2007efficient}, one can bound the size of terms in the expansion of the exponential at order $s$ by replacing each operator with its norm.
Replacing each operator in the exponentials of $S_{2k}$ by their norms, you have (corresponding to Eq.~(7) in \cite{berry2007efficient})
\begin{equation}\label{eq:powbnd}
    (1+|\lambda| \Lambda + (|\lambda| \Lambda)^2/2+\ldots)^{2L 5^{k-1}}.
\end{equation}
That gives the upper bound for the order-$s$ terms in $S_{2k}$ as
\begin{equation}\label{eq:orderserror}
    \frac{(2L 5^{k-1}|\lambda| \Lambda)^s}{s!}.
\end{equation}
This expression is specific to the Trotter-Suzuki product formulae.
Similarly, the order $s$ terms in the exact exponential of the Hamiltonian may be upper bounded as
\begin{equation}\label{eq:orderserror2}
    \frac{(L|\lambda| \Lambda)^s}{s!}.
\end{equation}
By summing Eq.~\eqref{eq:orderserror} and \eqref{eq:orderserror2}, and replacing $\lambda$ with $\lambda/2$, we can upper bound $\| D\|$ as
\begin{align}
    \| D \| &= \biggl\Vert \exp\biggl(\frac{\lambda}2 \sum_{j=1}^L H_j \biggr)-  S_{2k}  \biggr\Vert \nn
    &\leq \sum_{s=2k+1}^{\infty}\frac{(2L 5^{k-1}|\lambda| \Lambda/2)^s}{s!}+ \sum_{s=2k+1}^{\infty} \frac{(L|\lambda| \Lambda/2)^s}{s!}     \nn
    & \le \frac{(5^{k-1}|\lambda| L\Lambda)^{2k+1}}{(2k+1)!}\exp\left(5^{k-1}|\lambda| L\Lambda \right)
    +\frac{(|\lambda| L\Lambda/2)^{2k+1}}{(2k+1)!}\exp\left(|\lambda| L\Lambda/2 \right).
\end{align}

From the definition in Eq.~\eqref{A},
\begin{align}
    A &:= \sum_{l \in \gamma}\sum_{h=1}^{L_{l}} |\epsilon_{h,l}|,
\end{align}
which corresponds to the sum of the magnitudes of the terms in $V^\dagger D+DV^\dagger$, where both quantities are for $\lambda/2$.
To bound the norm of the higher-order terms in $V^\dagger D=V^\dagger S_{2k}-\openone$, we can consider the corresponding higher-order terms in $V^\dagger S_{2k}$.
Similarly, the higher-order terms in $DV^\dagger$ correspond to those in $S_{2k}V^\dagger$.

When multiplying $S_{2k}$ by the inverse of the evolution, one can use the same approach as for $\|D\|$, but the expression in \eqref{eq:powbnd} would be multiplied by
\begin{equation}
    (1+L|\lambda| \Lambda + (L|\lambda| \Lambda)^2/2+\ldots),
\end{equation}
for the exact exponential.
That is equivalent to replacing the power with $2L5^{k-1}+L$, so one can give the upper bound on the order-$s$ term as
\begin{equation}\label{eq:orderserrorVD}
    \frac{(2\times 5^{k-1}+1)^s(L|\lambda| \Lambda)^s}{s!}.
\end{equation}
Therefore, replacing $\lambda$ with $\lambda/2$, we can upper bound the size of the terms in $V^\dagger D+D V^\dagger$ by summing twice Eq.~\eqref{eq:orderserrorVD} to give
\begin{align}
    A \le 2\frac{[(5^{k-1}+1/2)|\lambda| L\Lambda]^{2k+1}}{(2k+1)!}\exp\left((5^{k-1}+1/2)|\lambda| L\Lambda \right) .
\end{align}
Note that it is trivially true that $\|D\|\le A/2$, because the sum of the magnitudes of the terms in $V^\dagger D+DV^\dagger$ upper bounds $\|V^\dagger D+DV^\dagger\|\ge 2\|D\|$.
\end{proof}
\end{document}